\documentclass[trackchanges]{aastex701}

\begin{document}

\title{Correlations between Dynamical Dark Energy and Neutrino Parameters}

\author[]{Ching-Yu Leung}
\affiliation{Department of Physics, The Chinese University of Hong Kong, Sha Tin, NT, Hong Kong}
\email[show]{ottoleung@link.cuhk.edu.hk}  

\author[]{WangZheng Zhang} 
\affiliation{Institut d'Astrophysique de Paris, UMR~7095, CNRS, Sorbonne Universit\'e, 98~bis~boulevard Arago, 75014~Paris, France}
\email[show]{wzzhang@link.cuhk.edu.hk}

\author[]{Ming-Chung Chu}
\affiliation{Department of Physics, The Chinese University of Hong Kong, Sha Tin, NT, Hong Kong}
\email[show]{mcchu@phy.cuhk.edu.hk}

\begin{abstract}
Recent Baryon Acoustic Oscillation (BAO) data from the Dark Energy Spectroscopic Instrument (DESI), fitted jointly with Cosmic Microwave Background (CMB) and Type Ia Supernovae (SNe) data, favor dynamical dark energy (DDE) over the standard $\Lambda$CDM model. In this work, we investigate whether the mass and degeneracy of relic neutrinos are degenerate with DDE parameters and can also account for this statistical deviation from $\Lambda$CDM. Using a Markov Chain Monte Carlo (MCMC) analysis of a joint Planck CMB temperature, DESI BAO, and Pantheon+ dataset, we show a robust parameter degeneracy between the DDE equation of state and neutrino physics. We supplement this with a semi-analytical framework that recovers the numerical correlation, showing that the tight constraint on the CMB acoustic scale drives this degeneracy. Our results highlight the clear correlation between DDE and neutrino parameters, underscoring the critical need to jointly constrain the neutrino sector and the dark sector in next-generation cosmological surveys.
\end{abstract}



\section{INTRODUCTION}
\label{Section1}
The Lambda cold dark matter ($\Lambda$CDM) model is one of the most successful cosmological models in describing the cosmic microwave background (CMB) anisotropy and large-scale structure \citep{aghanim2020planck,alam2021completed,abbott2022dark,balkenhol2023measurement, louis2025atacama, madhavacheril2024atacama}. However, recently, dynamical dark energy (DDE) models, such as the Chevallier-Polarski-Linder (CPL) parametrization of the dark energy (DE) equation of state (EoS), $P_\Lambda = [w_0 + (1-a)w_a]\rho_\Lambda$ \citep{chevallier2001accelerating, linder2003exploring},  received renewed attention \citep{gu2025dynamical, gialamas2025interpreting, giare2024robust, cortes2024interpreting} since deviations from the $\Lambda$CDM values of $(w_0,w_a)=(-1,0)$ at $2.8-4.2\sigma$ is found when fitted with Dark Energy Spectroscopic Instrument (DESI)'s latest baryon acoustic oscillation (BAO) data with CMB and different Type-IA supernovae (SNe) datasets \citep{abdul2025desi}, hinting at physics beyond $\Lambda$CDM. However, before interpreting such a deviation as evidence for DDE, it is important to examine whether DDE parameters could be degenerate with extensions in other cosmological sectors, which may explain the deviation. \\

A simple extension of $\Lambda$CDM model without introducing new physics is to allow finite neutrino masses and chemical potentials, namely the $\Lambda$CDM$\xi$ model \citep{yeung2021relic, yeung2025refitting}. Neutrino oscillation experiments provided a lower bound of the total neutrino mass $M_\nu \equiv \sum_i m_{\nu_i} > 0.058~\mathrm{eV}~(0.098~\mathrm{eV})$ for Normal (Inverted) Hierarchy, where $m_{\nu_i}$ is the $i$th mass eigenvalue \citep{esteban2024nufit}. We define neutrino degeneracy $\xi_\nu \equiv \mu_\nu/T_\nu$, where $\mu_\nu$ and $T_\nu$ are relic neutrino chemical potential and temperature, respectively. Although $\xi_{\nu_e}$ is tightly constrained to be $\ll 1$ by Big Bang Nucleosynthesis (BBN) data due to its direct impact on the weak interaction rates governing the neutron-to-proton ratio \citep{burns2023indications, escudero2023primordial, domcke2023new}, $\xi_{\nu_{\mu/\tau}} \sim \mathcal{O}(1)$ is still allowed \citep{yeung2021relic, zhao2025constraints, barenboim2017flavor}. Galaxy pairwise velocity measurement also suggests the degeneracy of the third neutrino mass eigenstate, $\xi_{\nu_3}\sim1$ \citep{zhang2026measuring}. Incorporating $M_\nu$ and $\xi_{\nu_3}$ in the $\Lambda$CDM model has been shown to successfully alleviate the $H_0$ and $S_8$ tensions when confronted with CMB temperature and lensing, BAO and weak lensing datasets (namely TTBW), showing a $\sim 3\sigma$ preference for both $M_\nu=0.57^{+0.17}_{-0.13}$ eV and $\xi_{\nu_3}=1.13^{+0.41}_{-0.19}$ \citep{yeung2025refitting} over Planck $\Lambda$CDM's values of $M_\nu=0.06$ eV and $\xi_{\nu_3}=0$, though the preference for non-zero $M_\nu$ and $\xi_{\nu_3}$ is weakened when CMB polarization data is included.\\

Both neutrino mass and chemical potential modify the neutrino energy density and hence the cosmic expansion history. Their effects on cosmological observables can therefore be degenerate with those induced by DDE. This raises the question of whether the apparent DE variation could partially be attributed to the effects of relic neutrinos instead. If so, characteristic correlations between ($M_\nu$, $\xi_{\nu_3}$) and DE parameters, ($w_0$, $w_a$), are expected when both sectors are allowed to vary simultaneously. We demonstrate below such a degeneracy and show the origin of the correlations between these parameters.\\

This paper is organized as follows. In Section~\ref{Section2}, we first describe the $\Lambda\mathrm{CDM}\xi$ and DDE models, and the setup for running  Markov Chain Monte Carlo (MCMC) analyses. We then present the MCMC result in Section~\ref{Section3}, which is studied with a semi-analytical approach detailed in Appendix~\ref{section6}. We summarize our findings in Section~\ref{Section4}.

\section{METHODOLOGY}
\label{Section2}
After decoupling, relic neutrinos remain in the mass eigenstates as time evolves. Following \cite{yeung2021relic}, we assume degenerate neutrino masses, i.e. $m_i=M_\nu/3$, where $i=1,2,3$, while $\xi_{\nu_i}$ in the mass eigenstate is related to $\xi_{\nu_\alpha}$ in the flavour eigenstate, where $\alpha=e,\mu,\tau$, by the Pontecorvo-Maki-Nakagawa-Sakata (PMNS) matrix \citep{barenboim2017flavor}. Given the strong mixing between $\nu_\mu$ and $\nu_\tau$, we assume $\xi_{\nu_\mu}=\xi_{\nu_\tau}$, and we fix $\xi_{\nu_e}=0.044$ as extracted from Big Bang Nucleosynthesis (BBN) data  \citep{burns2023indications}, leaving only one independent parameter describing neutrino degeneracies, which we choose to be $\xi_{\nu_3}$. The neutrino energy density becomes

\begin{equation}
    \rho_{\nu}(a)=\frac{1}{2\pi^2}\sum^3_{i=1}\int^\infty_0\left(\frac{\sqrt{p^2+(\frac{M_\nu}{3})^2}}{e^{\frac{pa}{T_\nu}-\xi_{\nu_i}}+1}+\frac{\sqrt{p^2+(\frac{M_\nu}{3})^2}}{e^{\frac{pa}{T_\nu}+\xi_{\nu_i}}+1}\right)p^2dp,
\end{equation}

\noindent which is characterized by $T_\nu$, $M_\nu$, and $\xi_{\nu_3}$ only, where $T_\nu\sim2~\mathrm{K}$ is the relic neutrino temperature today. The DE density with the CPL EoS is

\begin{equation}
    \rho_\Lambda(a)=\rho_{\Lambda0} a^{-3(1+w_0+w_a)}e^{3w_a(a-1)},
\end{equation}

\noindent where $\rho_{\Lambda0}$ is the dark energy density today. \\

In the following, we first demonstrate that $w_0w_a$CDM is still preferred over $\Lambda$CDM based on the specific combination of datasets (namely TTBS) we will use -- consisting of Planck CMB temperature, DESI BAO, and Pantheon+ SNe \citep{aghanim2020planck,abdul2025desi,brout2022pantheon+}. Similar to TTBW, we do not include CMB polarization data, since \cite{yeung2025refitting} found a possible inconsistency between the constraints inferred from CMB temperature and polarization data in the $\Lambda$CDM$\xi$ model. We therefore consistently adopt the CMB temperature-based dataset combination throughout this paper.\\

To probe the degeneracy between neutrino and DE parameters, we apply the Boltzmann code \textbf{CLASS} \citep{blas2011cosmic} with both massive neutrino and CPL DE activated. We call this combined model $w_0w_a\mathrm{CDM}\xi$ model in the following. Then, we perform MCMC fitting using \textbf{Cobaya} \citep{torrado2019cobaya,torrado2021cobaya} of the TTBS dataset with $R-1=0.01$ being the convergent criterion. We adopt Gaussian priors on six parameters $\Omega_b h^2, \Omega_c h^2, \tau_{reio}, \log(10^{10} A_\mathrm{s}), n_s, H_0$ as tabulated in Table~\ref{table:Gaussian} based on \cite{yeung2025refitting}'s result of the TTBW dataset, and flat priors on $w_0, w_a, M_\nu, \xi_{\nu_3}$ to probe the degeneracy around the parameter space favored by the CMB temperature constraint. Alongside MCMC fitting, we perform a semi-analytical analysis of the correlations among $w_0, w_a, M_\nu$ and $\xi_{\nu_3}$. In this approach, we assume that $w_0, w_a, M_\nu$ and $\xi_{\nu_3}$ are varied subject to fixed CMB acoustic scale and angle-averaged BAO angles. Any changes induced by one parameter should therefore be compensated by those induced by the rest of the parameters. Full details of the analysis are presented in Appendix~\ref{section6}.\\

\begin{table}[h]
\centering
\begin{tabular}{ccc}
\hline\hline
& Mean & Standard Deviation\\
\hline
$\Omega_b h^2$& 0.02325 & 0.00042\\
\hline
$\Omega_c h^2$ & 0.1223& 0.0051\\
\hline
$\tau_{reio}$ & 0.201& 0.045\\
\hline
$\log(10^{10} A_\mathrm{s})$ & 3.337& 0.096\\
\hline
$n_s$ & 1.019& 0.022\\
\hline
$H_0~[\mathrm{km~s^{-1}~Mpc^{-1}}]$ & 71.2& 2.1\\
\hline\hline
\end{tabular}
\caption{Mean and standard deviation values of Gaussian priors on six cosmological parameters $\Omega_b h^2, \Omega_c h^2, \tau_{reio}, \log(10^{10} A_\mathrm{s}), n_s, H_0$ in the MCMC fitting, adopted from \cite{yeung2025refitting}.}
\label{table:Gaussian}
\end{table}

\section{RESULTS}
\label{Section3}
The MCMC results are summarized in Table~\ref{table:MCMC}. Under this TTBS dataset, there are $2.9\sigma$ and $2.2\sigma$ deviations from $\Lambda$CDM's values for $w_0$ and $w_a$, respectively.\\

\begin{table}[h]
\centering
\begin{tabular}{cccccc}
\hline\hline
 & $w_0$ & $w_a$ & $M_\nu$ [{eV}] & $\xi_{\nu_3}$ & $H_0$ [$\mathrm{km\ s^{-1}\ Mpc^{-1}}$] \\
\hline
$w_0w_a\mathrm{CDM}$ & $-0.769\pm 0.079$ & $-0.68^{+0.31}_{-0.27}$  & $0.06$ & $0.00$ & $66.58\pm 0.79$\\
\hline
$w_0w_a\mathrm{CDM\xi}+\mathrm{prior}$ & $-0.796\pm0.079$ & $-0.56^{+0.34}_{-0.30}$ & $0.41\pm 0.13$ & $1.14^{+0.18}_{-0.13}$ & $69.7\pm 1.1$ \\
\hline\hline
\end{tabular}
\caption{Mean values of $w_0, w_a, M_\nu, \xi_{\nu_3}, H_0$ (68\% CL) for $w_0w_a\mathrm{CDM}$ and $w_0w_a\mathrm{CDM}\xi$ with Gaussian priors obtained from MCMC fitting to the TTBS dataset.}
\label{table:MCMC}
\end{table}

\begin{figure}[h!]
    \centering
    \includegraphics[width=1.0\linewidth]{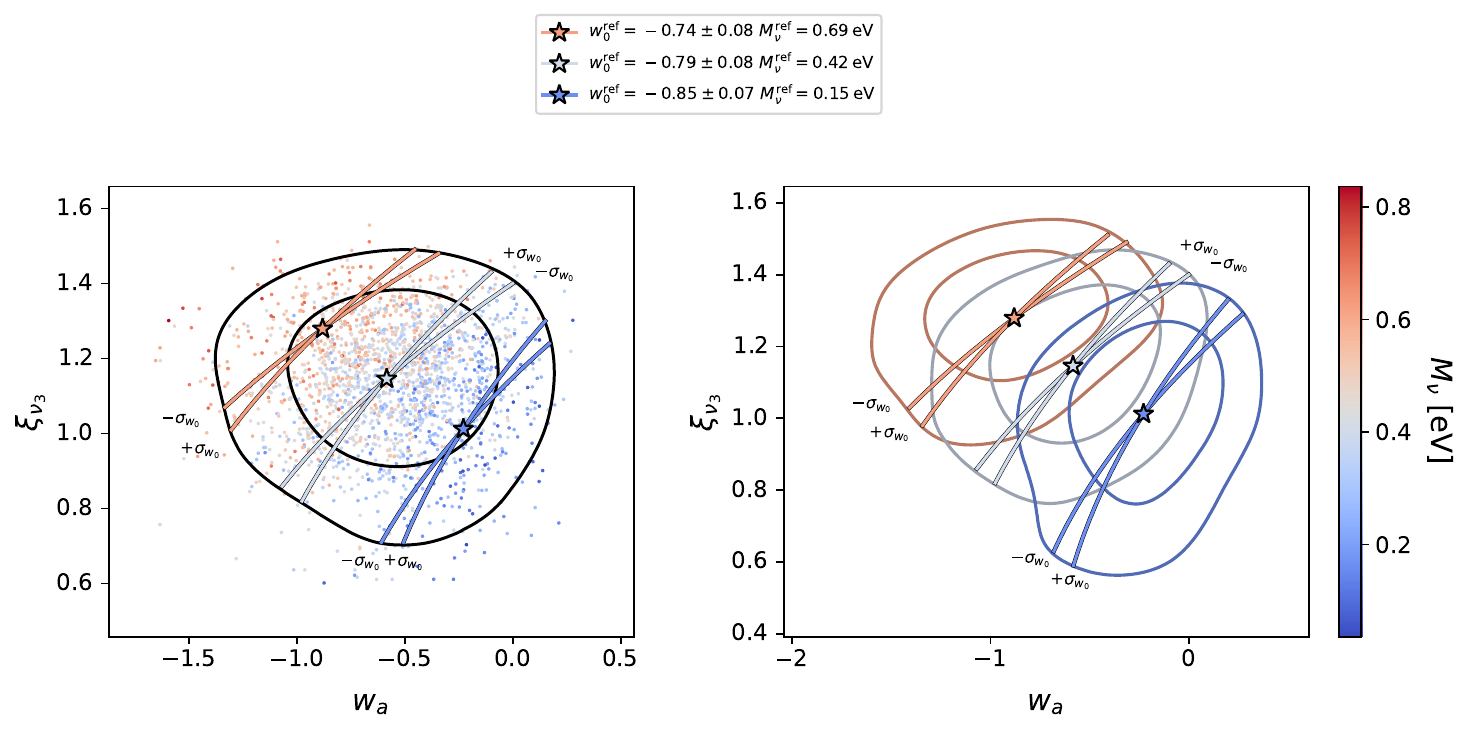}
    \caption{Left Panel: 2D contours ($68\%$ and $95\%$ CL) for $w_a$ and $\xi_{\nu_3}$ obtained from the whole MCMC sample of fitting $w_0w_a\Lambda\mathrm{CDM}\xi$+prior to the TTBS dataset. Scatter points are colored based on values of $M_\nu$, as indicated by the color bar. Right Panel: 2D contours ($68\%$ and $95\%$ CL) for $w_a$ and $\xi_{\nu_3}$ obtained from the three MCMC sub-samples grouped by $M_\nu$, colored based on the mean value of $M_\nu$ in each group. In both panels, the six curves represent Eq.~\ref{eq:perturbation} expanded at different reference points denoted by the stars, corresponding to different rows in Table~\ref{table:perturbation}.}
    \label{fig:MCMC}
\end{figure}

We perform a semi-analytical analysis of the DDE and neutrino parameters \citep{yeung2021relic, yeung2025refitting} (see Appendix~\ref{section6} for details). We assume that the well-determined CMB acoustic scale and angle-averaged BAO angles are fixed, and we vary the DDE and neutrino parameters from their reference values, $(w_0^{\rm ref}, w_a^{\rm ref}, M_\nu^{\rm ref}, \xi_{\nu_3}^{\rm ref})$, to obtain Eq.~\ref{eq:perturbation}. 
\begin{equation}
0
= C_{w_a}(w_a-w^{\rm ref}_a) + C^{(1)}_{\xi_{\nu_3}}(\xi_{\nu_3}-\xi_{\nu_3}^{\rm ref}) + C^{(2)}_{\xi_{\nu_3}}(\xi_{\nu_3}-\xi_{\nu_3}^{\rm ref})^2,
\label{eq:perturbation}
\end{equation}

\noindent where the upper indices in the parenthesis of $C_{\xi_{\nu_3}}$ refer to their orders. We focus on three sub-samples grouped by $M_\nu$: $0.05~\mathrm{eV}\leq M_\nu \leq 0.2~\mathrm{eV},~0.35~\mathrm{eV}\leq M_\nu \leq 0.5~\mathrm{eV}$, and $ 0.65~\mathrm{eV}\leq M_\nu \leq 0.8~\mathrm{eV}$, and then we extract the mean of ${w}_a,\xi_{\nu_3}$, and $M_\nu$ from the MCMC data points of $w_0w_a\Lambda\mathrm{CDM}\xi$+prior fitting of each group. These are the reference points $w_a^{\rm ref}, M_\nu^{\rm ref},$ and $\xi_{\nu_3}^{\rm ref}$, respectively. We also take the mean $\pm$1 SD to be the reference points of $w_0$ for each group. These reference points are listed in Table~\ref{table:perturbation}. We then obtain the coefficients $C_{w_a}, C^{(1)}_{\xi_{\nu_3}}$ and $C^{(2)}_{\xi_{\nu_3}}$ corresponding to the six sets of reference points. These are all summarized in Table~\ref{table:perturbation} with each row corresponding to a perturbation curve via Eq.~\ref{eq:perturbation} plotted in Figure~\ref{fig:MCMC}. All other cosmological parameters except 
$w_0, w_a, M_\nu$ and $\xi_{\nu_3}$ are fixed at the mean values of the entire MCMC sample.

\begin{table}[h]
\centering
\begin{tabular}{cccc|ccc}
\hline\hline
$w_0^{\mathrm{ref}} (\bar{w}_0\pm\sigma_{w_0})$ & $M_\nu^{\mathrm{ref}}~\mathrm{[eV]}$ & $w_a^{\mathrm{ref}}$ & $\xi_{\nu_3}^{\mathrm{ref}}$ & $C_{w_a}$ & $C^{(1)}_{\xi_{\nu_3}}$ & $C^{(2)}_{\xi_{\nu_3}}$ \\
\hline
$-0.74+0.08$& $0.69$ & $-0.88$ & $1.28$ & $-2.12$ & $3.86$ & $1.99$\\
\hline
$-0.74-0.08$& $0.69$ & $-0.88$ & $1.28$ & $-1.61$ & $3.90$ & $2.00$ \\
\hline
$-0.79+0.08$&$0.42$ & $-0.58$ & $1.15$ & $-2.47$ & $3.63$ & $2.01$ \\
\hline
$-0.79-0.08$&$0.42$ & $-0.58$ & $1.15$ & $-1.82$ & $3.67$ & $2.03$ \\
\hline
$-0.85+0.07$& $0.15$ & $-0.23$ & $1.01$ & $-3.00$ & $3.37$ & $2.03$ \\
\hline
$-0.85-0.07$& $0.15$ & $-0.23$ & $1.01$& $-2.22$ & $3.41$ & $2.06$ \\
\hline\hline
\end{tabular}
\caption{Reference values of $w_0, w_a, M_\nu$ and $\xi_{\nu_3}$, and their corresponding coefficients $C_{w_a}, C^{(1)}_{\xi_{\nu_3}}$ and $C^{(2)}_{\xi_{\nu_3}}$ in the perturbation equation (Eq.~\ref{eq:perturbation}).}
\label{table:perturbation}
\end{table}

\begin{figure}[h!]
    \centering
    \includegraphics[width=0.7\linewidth]{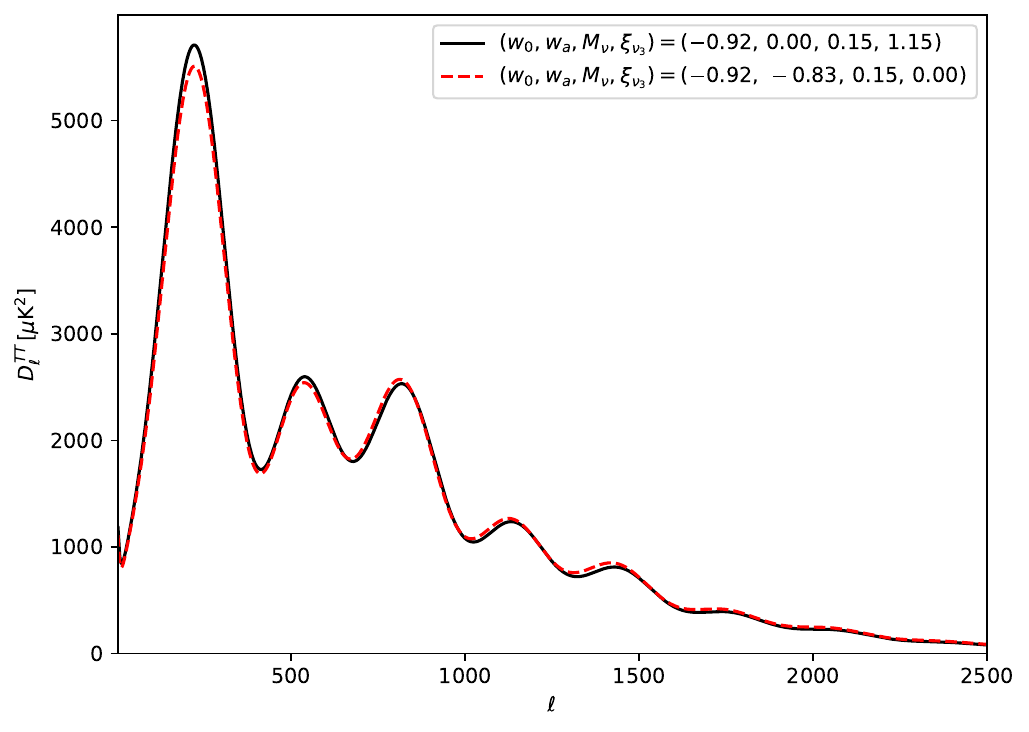}
    \caption{CMB temperature power spectra $D_\ell^{TT}=\ell(\ell+1)C_\ell^{TT}/2\pi$ with $\omega_a = 0~(\xi_{\nu3} = 0$) and $\xi_{\nu3} = 1.15 ~(\omega_a = -0.83),$ shown as black solid (red dashed) line.}
    \label{fig:PowerSpectrum}
\end{figure}

As shown in Figure~\ref{fig:MCMC}, curves based on Eq.~\ref{eq:perturbation} align well with the MCMC result for fixed $M_\nu$, suggesting the correlations between the neutrino and DE parameters arises from CMB acoustic scale and BAO angles, with the former dominating due to a significantly smaller error bar. Applying the perturbation curve described by Row 6 in Table~\ref{table:perturbation}, we calculated the corresponding $\xi_{\nu_3}$ (and $w_a$) by assuming $w_a=0$ (and $\xi_{\nu_3}=0$) while keeping $w_0$ and $M_\nu$ at their reference values. Using these parameters, we show in Fig.~\ref{fig:PowerSpectrum} that the corresponding CMB temperature power spectra overlap very well, particularly their peak and trough positions. This confirms that the two cosmological models, which fall on the same perturbation curve, have the same CMB acoustic scale to a good approximation. In Figure~\ref{fig:MCMC}, we see a clear correlation between $w_a$ and $\xi_{\nu_3}$ for fixed $w_0$ and $M_\nu$. Furthermore, for the same $M_\nu$, the curve with $+\sigma_{w_0}$ and hence a larger $w_0$ is steeper. Increasing $w_0$ and $M_\nu$ together shift these curves upward, demanding a larger $\xi_{\nu_3}$. Overall, the correlations among neutrino and DE parameters are clearly shown.\\

\section{CONCLUSIONS}
\label{Section4}
In this paper, we demonstrate the correlations between DDE parameters, $w_0$ and $w_a$, and neutrino parameters, $M_\nu$ and $\xi_{\nu_3}$, using MCMC fitting to a combination of datasets, including Planck CMB temperature, DESI BAO, and Pantheon+ SNe. We also present a semi-analytical analysis, showing that the dominant cause of the correlations is the well-determined CMB acoustic scale. The result implies a degeneracy between the DDE and $\Lambda$CDM$\xi$ models, at least with respect to CMB and BAO acoustic scales. As such, introducing neutrino parameters together with DDE faces severe penalties due to redundancy, which makes it statistically less favorable. To distinguish the two models, we need tight measurements that constrain cosmological models beyond the expansion history. Precise measurements on the growth of large-scale structure could be the key.\\

\section{ACKNOWLEDGEMENTS}
\label{section5}
All computational resources used in this paper were provided by the Chinese University of Hong Kong Central Research Computing Cluster. This research is supported by grants from the Research Grants Council of the Hong Kong Special Administrative Region, China, under Projects AoE/P-404/18 and 14300223. Besides, \textbf{Getdist} \citep{Lewis:2019xzd} and \textbf{Matplotlib} \citep{Hunter:2007} were used to visualize data while \textbf{Scipy} \citep{2020SciPy-NMeth} and \textbf{Numpy} \citep{harris2020array} were used to perform calculations for the semi-analytical analyses.

\appendix{}
\label{appendix}

\section{Semi-Analytical Study of Correlation}
\label{section6}
We perform a semi-analytical analysis following \cite{yeung2021relic, yeung2025refitting} on the inverse of CMB acoustic scale $1/\theta_*$ and the angle-averaged BAO angle $\beta_{\rm avg}(z)$ to study the correlations between different parameters. In general, we obtain a the relation between all four parameters $w_0,w_a,M_\nu,\xi_{\nu_3}$ together:
\begin{equation}
0
= C_{w_0}(w_0-w^{\rm ref}_0) + C_{w_a}(w_a-w^{\rm ref}_a)+ C_{M_\nu}\frac{M_\nu-M_\nu^{\rm ref}}{M_\nu^{\rm ref}} + C^{(1)}_{\xi_{\nu_3}}(\xi_{\nu_3}-\xi_{\nu_3}^{\rm ref}) + C^{(2)}_{\xi_{\nu_3}}(\xi_{\nu_3}-\xi_{\nu_3}^{\rm ref})^2,
\end{equation}
where $C_{w_0,w_a,M_\nu},C^{(1)}_{\xi_{\nu_3}}$ and $C^{(2)}_{\xi_{\nu_3}}$ are coefficients derived below. However, in practice, we only visualize each pair of them in a contour plot such as Figure~\ref{fig:MCMC}, fixing the other two to their reference values.
Since we have more than one measurements, including multiple BAO angles and one CMB acoustic scale, the weighted average is taken at the end.

\subsection{Perturbation on $1/\theta_*$}
The inverse of the CMB acoustic scale $1/\theta_*$ is defined by
\begin{equation} 
\frac{1}{\theta_{*}}
= \frac{D_M(z_*)}{r_*},
\end{equation}
where $r_*$ is the comoving sound horizon of the photon-baryon fluid at the photon decoupling epoch $z_*$ with sound speed $c_s$, and $D_M$ is the transverse comoving distance:
\begin{equation} 
r_{*}
=\int^{\infty}_{z_{*}}\frac{c_s(z')dz'}{H(z')}, 
\end{equation}
\begin{equation} 
D_M(z)
= \int_0^z \frac{dz'}{H(z')}. 
\end{equation} 
We assume $1/\theta_*$ is fixed, i.e.
\begin{equation}
\delta \left(\frac{1}{\theta_*}\right)
= 0.
\end{equation}
Since we focus on the local relation around the reference point, Taylor expansion is taken up to the first order for $w_0,w_a,M_\nu$ and second order for $\xi_{\nu_3}$ (the first order of $\xi_{\nu_3}$ is negligible at some reference points where $\xi_{\nu_3}\sim0$), which yields the following:
\begin{equation}
C^*_{w_0}
=I_1^*[K_{w_0}(z_*)],\end{equation} \begin{equation} 
C^*_{w_a}
=I_1^*[K_{w_a}(z_*)],\end{equation} \begin{equation} 
C^*_{M_\nu}
=M^{\rm ref}_{\nu}I_1^*[K_{M_\nu}(z_*)],\end{equation} \begin{equation} 
C^{*(1)}_{\xi_{\nu_3}}
=I_1^*[K_{\xi_{\nu_3}}(z_*)],
\end{equation} 
\begin{equation} 
C^{*(2)}_{\xi_{\nu_3}}
= \frac{1}{2}\frac{\partial C^{(1)}_{\xi_{\nu_3}}}{\partial \xi_{\nu_3}}, \end{equation}
where the upper indices in the parenthesis of $C^*_{\xi_{\nu_3}}$ denote their orders, and
\begin{equation}
I_1^*[f(z)]
\equiv\frac{D_M(z)}{r_{*}^2}\int_z^\infty dz'c_s(z')f(z')-\frac{1}{r_{*}}\int_{0}^{z} dz'f(z'),\end{equation}
\small
\begin{equation} 
K_{w_0}(z)
\equiv
\frac{H_0^2}{2H^3(z)}
3\ln(1+z)\Omega_\Lambda e^{3(1+w_0+w_a)\ln(1+z)+3w_a(\frac{1}{1+z}-1)} ,\end{equation} \begin{equation} 
K_{w_a}(z)
\equiv
\frac{H_0^2}{2H^3(z)}
3(\ln(1+z)+\frac{1}{1+z}-1)\Omega_\Lambda e^{3(1+w_0+w_a)\ln(1+z)+3w_a(\frac{1}{1+z}-1)} , \end{equation} \begin{equation} 
K_{M_\nu}(z)
\equiv
\frac{H_0^2}{2H^3(z)} \sum^3_{i=1} \frac{1}{2\pi^2} \int_0^\infty 
\frac{M_\nu}{9\sqrt{p^2+\left(\frac{M_\nu}{3}\right)^2}} p^2\left(
\frac{1}{e^{p/((1+z)T_{\nu})+\xi_{\nu_i}}+1}
+
\frac{1}{e^{p/((1+z)T_{\nu})-\xi_{\nu_i}}+1}
\right)dp, \end{equation} \begin{equation} 
K_{\xi_{\nu_3}}(z)
\equiv
\frac{H_0^2}{2H^3(z)}
\sum^3_{i=1} \frac{1}{2\pi^2}
\int_0^\infty \sqrt{p^2+\left(\frac{M_\nu}{3}\right)^2} p^2\frac{d\xi_{\nu_i}}{d\xi_{\nu_3}}
\left(
\frac{-e^{p/((1+z)T_{\nu})+\xi_{\nu_i}}}{(e^{p/((1+z)T_{\nu})+\xi_{\nu_i}}+1)^2}
+
\frac{e^{p/((1+z)T_{\nu})-\xi_{\nu_i}}}{(e^{p/((1+z)T_{\nu})-\xi_{\nu_i}}+1)^2}
\right)dp. \end{equation}

\subsection{Perturbation on $\beta_{\rm avg}(z)$}
\label{A2}
The angle-averaged BAO angle detected at redshift $z$ is given by
\begin{equation} 
\beta_{\rm avg}(z)
= \left(z \beta_\perp^2(z) \beta_\parallel(z)\right)^{1/3}, 
\end{equation}
where $\beta_\perp(z)$ and $\beta_\parallel(z)$ are BAO angles perpendicular and parallel to the line of sight
\begin{equation}
\beta_\perp(z)
= \frac{D_M(z)}{r_d},
\end{equation}
\begin{equation}
\beta_\parallel(z)
= \frac{1}{r_{d} H(z)}.
\end{equation}
$r_d$ is the comoving sound horizon of the photon-baryon fluid at the drag epoch $z_d$:
\begin{equation} 
r_{d}
=\int^{\infty}_{z_{d}}\frac{c_s(z')dz'}{H(z')}. 
\end{equation} 
Similarly, we assume
\begin{equation}
\delta \beta_{\rm avg}(z)
= 0,
\end{equation}
which yields the following:
\begin{equation}
C^d_{w_0}
=\frac{z\beta_{\perp}(z)}{3\beta_{\rm avg}^{2}(z)}\left(2\beta_\parallel(z)I_1^d[K_{w_0}(z)]+\beta_\perp(z)I_2^d[K_{w_0}(z)]\right),\end{equation} \begin{equation} 
C^d_{w_a}
=\frac{z\beta_{\perp}(z)}{3\beta_{\rm avg}^{2}(z)}\left(2\beta_\parallel(z)I_1^d[K_{w_a}(z)]+\beta_\perp(z)I_2^d[K_{w_a}(z)]\right),\end{equation} \begin{equation} 
C^d_{M_\nu}
=M^{\rm ref}_{\nu}\frac{z\beta_{\perp}(z)}{3\beta_{\rm avg}^{2}(z)}\left(2\beta_\parallel(z)I_1^d[K_{M_\nu}(z)]+\beta_\perp(z)I_2^d[K_{M_\nu}(z)]\right),\end{equation} \begin{equation} 
C^{d(1)}_{\xi_{\nu_3}}
=\frac{z\beta_{\perp}(z)}{3\beta_{\rm avg}^{2}(z)}\left(2\beta_\parallel(z)I_1^d[K_{\xi_{\nu_3}}(z)]+\beta_\perp(z)I_2^d[K_{\xi_{\nu_3}}(z)]\right),
\end{equation} 
\begin{equation} 
C^{d(2)}_{\xi_{\nu_3}}
= \frac{1}{2}\frac{\partial C^{d(1)}_{\xi_{\nu_3}}}{\partial \xi_{\nu_3}},
\end{equation} 
\small
where we define
\begin{equation}
I_1^d[f(z)]
\equiv\frac{D_M(z)}{r_{d}^2}\int_{z_{d}}^\infty dz'c_s(z')f(z')-\frac{1}{r_{d}}\int_{0}^z dz'f(z'),\end{equation}
\begin{equation}
I_2^d[f(z)]
\equiv\frac{1}{r_{d}^2H(z)}\int_{z_{d}}^\infty dz'c_s(z')f(z')-\frac{f(z)}{r_{d}}.
\end{equation}
\normalsize
\subsection{Weighted Average of Coefficients}
Each of the CMB acoustic scale and BAO angles at different redshifts gives a set of different values of the coefficients. We take the weighted average of each coefficient according to
\begin{equation}
\bar{C}
=
\frac{
    W_{\mathrm{CMB}}\, C_{\mathrm{CMB}}
    + \displaystyle\sum_{i=1}^{N_{\mathrm{BAO}}}
      W_{\mathrm{BAO},i}\, C_{\mathrm{BAO},i}
}{
    W_{\mathrm{CMB}}
    + \displaystyle\sum_{i=1}^{N_{\mathrm{BAO}}}
      W_{\mathrm{BAO},i}
},
\label{eq:weighted-mean-C}
\end{equation}

\noindent where $N_{\mathrm{BAO}}$ is the number of BAO measurements, $W$ is the weighting defined by $W=1/\sigma^2$, and $\sigma$ is the error bar of each measurement. Note that we apply the same perturbation analysis to different kind of BAO angles, $\beta_{\rm avg}(z), \beta_\perp(z)$, and $\beta_\parallel(z)$, according to \textbf{Appendix}~\ref{A2}.


\bibliography{sample701}{}
\bibliographystyle{aasjournalv7}



\end{document}